\documentstyle[aps,preprint]{revtex}
\begin{document}
\draft

\title{Quantum Entanglement and Entropy}
\author{Filippo Giraldi, Paolo Grigolini$^{1,2,3}$}
\address{$^{1}$Dipartimento di Fisica dell'Universit\`{a} di Pisa and
INFM,
Piazza Torricelli 2, 56127 Pisa, Italy}
\address{$^{2}$Istituto di Biofisica CNR, Area della Ricerca di Pisa,
Via Alfieri 1, San Cataldo 56010 Ghezzano-Pisa,  Italy}
\address{$^{3}$Center for Nonlinear Science, University of North Texas,\\
P.O. Box 5368, Denton, Texas 76}
\date{\today}
\maketitle

\begin{abstract}
Entanglement is the fundamental quantum property behind the now
popular field of quantum transport of information. This quantum
property is incompatible with the separation of a single system
into two uncorrelated subsystems. Consequently, it  does not
require the use of an additive form of entropy. We discuss the
problem of the choice of the most convenient entropy indicator,
focusing our attention on a system of 2 qubits, and on a special
set, denoted by $\Im$. This set contains both the maximally and
the partially entangled states that are described by density
matrices diagonal in the Bell basis set. We select this set for
the main purpose of making more straightforward our work of
analysis. As a matter of fact, we find that in general the
conventional von Neumann entropy is not a monotonic function of
the entanglement strength. This means that the von Neumann entropy
is not a reliable indicator of the departure from the condition of
maximum entanglement. We study the behavior of a form of
non-additive entropy, made popular by the 1988 work by Tsallis. We
show that in the set $\Im$, implying the key condition of
non-vanishing entanglement, this non-additive entropy indicator
turns out to be a strictly monotonic function of the strength of
the entanglement, if entropy indexes $q$ larger than a critical
value $Q$ are adopted. We argue that this might be a consequence
of the non-additive nature of the Tsallis entropy, implying that
the world is quantum and that uncorrelated subsystems do not
exist.

\end{abstract}


\section{introduction}
The entanglement is the fundamental quantum property behind the
interesting process of quantum teleportation proposed some years
ago by Bennett \emph{et al}. \cite{bennett}. For this reason it is
important to quantify entanglement \cite{hill}. It has also been
found that the fidelity of the quantum teleportation is always
larger than that of any classical communication protocol even in
the noisy environment\cite{popescu}. However, for the
teleportation to take place perfectly, it is necessary for them to
share a maximally entangled state. This generates the need for
special purifications protocols.  On the other hand, the
statistical analysis of these protocols shares deep similarities
with the second principle of
thermodynamics\cite{plenio,thermodynamics1,thermodynamics2}.

In spite of the plausible conjecture that there exists a deep
connection between quantum teleportation and
thermodynamics\cite{plenio,thermodynamics1,thermodynamics2}, the
entanglement is expressed by means of an entropic structure, the
conventional von Neumann entropy, only in the case of pure states.
In general the definitions of entanglement dictated by the
purification protocol are not directly related to entropy
indicators. Cerf and Adami\cite{conjecture} show that the
conditional von Neumann entropy can become negative, thereby
pointing out the non-ordinary information aspects of quantum
entanglement. However, the reason for the lack of a direct
connection between quantum entanglement of mixed states and
entropy indicator is probably that, as shown in this paper, the
von Neumann entropy is not a reliable indicator of
entanglement\cite{conjecture}. The inadequacy of the Shannon
information, and consequently of the von Neumann entropy as an
appropriate quantum generalization of Shannon
entropy\cite{notebrukner}, has already been pointed out by Brukner
and Zeilinger\cite{attempt5,brukner}. It is notable that according
to these authors a new kind of entropy indicator is made necessary
by the following deep difference between quantum and classical
information. In classical measurement the Shannon information is a
natural measure of our ignorance about property of a system, whose
existence is independent of measurement. Quantum measurement, on
the contrary, cannot be claimed to reveal system's properties
existing before the measurement is done.

In this paper we focus on the earlier mentioned quantum property,
essential for teleportation of information: entanglement.
Entanglement implies that a system cannot be divided into two
uncorrelated subsystems, and this, in turn, makes useless the
ordinary request for an additive form of entropy.  Thus, in this
paper we explore and discuss the possible benefits stemming from
the adoption of the non-additive entropy indicator advocated years
ago by Tsallis\cite{originalpaper}. The Tsallis entropy is applied
to a large number of physical conditions, characterized by the
existence of extended correlation\cite{brazil}. The use of this
form of non-additive indicator in the field of quantum
teleportation, to the best of our knowledge, has been discussed in
only a few papers\cite{attempt1,attempt2,attempt3,attempt4}. The
paper of Ref.\cite{attempt1}
    claims to the realization of a greater sensitivity to the occurence
    of a dephasing process resulting in the annihilation of any form of
    entanglement.
The papers of Refs. \cite{attempt2,attempt3,attempt4} point out
the efficiency of Tsallis entropy for the detection of the
breakdown of local realism. The paper of Ref.\cite{attempt3} aims
at proving that the Jaynes principle\cite{jaynes1,jaynes2},
applied to the non-extensive entropy, yields naturally the
entangled state. The present paper, although based on the adoption
of the Tsallis non-extensive entropy as the paper of Ref.
\cite{attempt3}, adopts a quite different perspective, which does
not rest on the adoption of the Jaynes principle.

The outline of the paper is as follows. Section II is devoted to a
concise illustration of the main properties of the non-extensive
entropy at work in this paper. Section III, devoted to the
entanglement of formation, shows that the von Neumann entropy, in
general, is not a monotonic function of entanglement, while the
non-extensive entropy is a monotonically increasing function of
the entanglement for suitably large values of the entropy index.
In Section IV we study the entanglement of a de-phasing process,
and we extend the monotonic properties of non-extensive entropy to
a condition more general than that of Section III. Section IV is
devoted to concluding remarks.

\section{The non-extensive entropy}
The entropy indicator applied in this paper has the form
\begin{equation}
  S_{q}(\rho) \equiv Tr \frac{\rho -\rho^{q}}{q-1}.
  \label{tsallis}
  \end{equation}
This form has been originally proposed by
Tsallis\cite{originalpaper} for the purpose of establishing the
most convenient thermodynamic perspective for fractal processes.
It is worth remarking that this is a generalization of the
conventional Gibbs-Shannon entropy indicator, whose explicit form
is recovered from Eq. (\ref{tsallis}) in the limiting case $q
\rightarrow 1$. This entropy indicator does not fit the additivity
condition, namely, the requirement that the entropy of a system $A
+ B$, consisting of two statistically  independent subsystems, $A$
and $B$, be the sum of the entropies of these two subsystems. In
fact, the definition of Eq. (\ref{tsallis}) yields, in this case,
the following equality
\begin{equation}
     S_{q}(A + B) = S_{q}(A ) + S_{q}( B) + (1-q) S_{q}(A ) S_{q}(B),
     \label{breakdownofthe additiveproperty}
     \end{equation}
making it evident that the additive property is recovered only in
the case $q = 1$, which, as earlier noted, makes the non-extensive
entropy of Eq. (\ref{tsallis}) become equivalent to the usual
Shannon entropy.

According to Tsallis and to the advocates of this non-extensive entropy
indicator (for a review, see Ref.\cite{brazil}), the violation of the
additive condition turns into a benefit when this
     entropy indicator is used to the study cases where the
     ideal condition of statistical independence is prevented by the
     nature itself of the processes under study\cite{brazil}. Notable
     examples are the processes with the long-range
     correlation\cite{brazil}. We think that quantum entanglement, which
     is the basic property for teleportation\cite{bennett}, is probably
     the most evident example of a condition incompatible with
the existence of uncorrelated subsystems. For this reason, the additivity
condition can be safely renounced, and the adoption of an entropy
index $q \neq 1$ might turn out to be beneficial.
    This paper is devoted to discussing to what an extent this
    conjecture proves to be correct.

\section{The Tsallis entropy at work: Entanglement of formation}
In this section we derive the central result of this paper. We
show that in the case of initial and final states, both with
non-vanishing entanglement, and described by a statistical density
matrix diagonal in the Bell basis, the Tsallis entropy decreases
upon increase of the entanglement of formation. The results of
this section refer to the entanglement of formation
\cite{wootters} of a system of two qubits. Thus, to make this
paper as self-contained as possible, we give here a concise
illustration of this key measure of entanglement. The main point
is that the entanglement property is defined without any ambiguity
for pure states\cite{bennett}. The entanglement of formation
extends to the statistical case the ordinary definition as
follows. Let us denote with  $\rho$ the statistical density matrix
for the mixed state of a space $ S_{1/2}^{(1)}\times S_{1/2
}^{(2)} $ of two spin-1/2 particles. The \emph{entanglement of
formation} \cite{bw}, denoted by the symbol $E_{F}$ throughout, is
defined as the minimum average entanglement of every ensemble of
pure states that represents $\rho$:
\begin{equation}
E_{F} \left( \rho \right)= \min_{\rho=\sum_{i} P_{i}
|\alpha_{i}\rangle\langle\alpha_{i}|} \sum_{i} P_{i} E\left(|
\alpha_{i} \rangle\right),
\end{equation} where $E(| \alpha_{i} \rangle)$ denotes in fact
the entanglement of a pure state, which is defined according to the
usual prescription\cite{bennett} by the expression
\begin {equation}
E\left(|\alpha\rangle \right)= - Tr \left(\rho_{A} \log_{2}
\rho_{A} \right)= - Tr \left(\rho_{B} \log_{2} \rho_{B} \right)
\end {equation}with
\begin{equation}
\rho_{A}\equiv Tr_{2} \rho,\hspace{1em}
\rho_{B}\equiv Tr_{1} \rho.
\end{equation}

Wootters, in an enlightening article\cite{wootters}, derives
an explicit formula for the entanglement of formation
of any arbitrary mixed state of a system of 2 qubits. This formula
reads:
\begin{equation}
E_{F}\left(\rho\right)=h\left(\frac{1+\sqrt{1-C^{2}\left(\rho\right)}}{2}
\right),
\end{equation}
where
\begin{equation}
h(x)\equiv -x\log_{2}x-(1-x)\log_ {2}{(1-x)}  .
\end{equation}
The quantity
$C\big(\rho\big)$, referred to by Wootters as \emph{concurrence}, is
defined by
\begin{equation}
C\left(\rho\right) \equiv \max
\Big\{0,\lambda_{m}-\lambda_{1}-\lambda_{2}-\lambda_{3} \Big\},
\end{equation}
where $\lambda_{m}$, $ \lambda_{1}$, $\lambda_{2}$ and $\lambda_{3}$
are the square roots of eigenvalues of the matrix
$\rho\cdot\widetilde{\rho}$, set in a decreasing order, with
$\lambda_{m}$ being the maximum eigenvalue. The matrix
$\tilde{\rho}$ denotes the spin-flipped
state:
     \begin{equation}
     \tilde{\rho}
\equiv
\left(\sigma_{y}\otimes\sigma_{y}\right)\rho^{*}
\left(\sigma_{y}\otimes\sigma_{y}\right).
\label{spinflip}
\end{equation}
The values of the entanglement of formation range from $0$ to $1$.
Furthermore, $E_{F}$ is a monotonic increasing function of $C$.
The values of the concurrence $C$, in turn, range from $0$ to $1$,
as the values of $E_{F}$ do. Consequently, the concurrence itself
can be considered as a proper measure of the entanglement of
formation. Note that the entanglement of formation is the only
kind of entanglement studied in the present paper. For simplicity,
we shall often refer to it simply as entanglement.

\subsection{On $\Im$, the working basis set for a system of 2 qubits}
The most general expression \cite{Horodecki1} of a mixed state $\rho$ of
the space $S_{1/2}^{(1)}\otimes S_{1/2}^{(2)}$ is:
\begin{equation}
\rho=\frac{1}{16}\left\{1\otimes 1+ \left(\sum_{i=1}^{3} r_{i}
     \sigma_{i}\right)\otimes 1+1\otimes \left( \sum_{i=1}^{3} s_{i}
     \sigma_{i}\right) + \sum_{i=1}^{3}\sum_{j=1}^{3} t_{ij}
   \sigma_{i}\otimes\sigma_{j}\right\},
   \end{equation}
    where $r_{i}, s_{i}$ and $t_{ij}$ are real parameters.
     Due to the excessive number of involved parameters, the direct use
     of
     this expression would make the calculation too complicated. For
     this reason we decided to limit our investigation to the set
     defined by:
\begin{equation}
\Im \equiv \left\{\rho : C(\rho)=2 P_{m}-1
> 0 \right\}, \label{working set}
\end{equation}
where $P_{m}$ denotes the greatest eigenvalue of the density
matrix $\rho$ describing the quantum state. On behalf of future
developments, we remark that the condition $P_{m}> 1/2$ makes this
maximum value unique. The adoption of the set $\Im$, as our
working set, does not rule out the possibility of considering
physical conditions of interest for the field of quantum
teleportation. In fact, the set $\Im$ contains the  states with
positive entanglement of formation $\left(E_{F}\left(\rho
\right)>0 \right)$ that are described by density matrices
corresponding to the solution of the equation
\begin{equation}
\rho\cdot\widetilde{\rho}=\rho^{2}.
\end{equation}
These density matrices, recently used by Bennett and
co-workers\cite{bw} (see also\cite{bw2}), become diagonal when
expressed in the Bell basis set. The set $\Im$ contains the
entangled Werner states\cite{wernerstate} as well as the maximally
entangled states. We also note that any mixed state can be brought
into a diagonal form in the Bell basis set by random bilateral
rotations\cite{bw}. This implies that, in spite of the
simplification made, our discussion is still rather general. The
problem here under discussion is the significance of the Tsallis
entropy as a measure of entanglement. The working set of states
that we select, $\Im$, does not conflict with the possibility of
discussing this issue with reference to one of the relevant
physical conditions that have been recently examined by the
investigators of this field of research\cite{bw,bw2}. In fact, we
shall study the change of the Tsallis entropy indicator upon the
entanglement change of a given pair of particles. This change
might be the result of a purification process such as that studied
by Bennett and co-workers\cite{bw,bw2}. These authors have
discussed the purification of a couple of Werner states, showing
how to increase the entanglement of formation of a pair at the
price of decreasing that of the other. Here, we discuss the
problem of detecting this increased entanglement with the Tsallis
entropy. We found to be relatively easy to establish this
important result within the set $\Im$. However, we cannot rule out
the possibility that this property is shared by any other set of
states resulting in the concurrence being an increasing function
of $P_{m}$. In all these cases  we might find the same property of
the entropy being proportional to the inverse of the entanglement.

It is worth to anticipate an aspect of fundamental importance. The
plausible reason why, as we shall see, the inverse of the Tsallis
entropy indicator is a successful measure of entanglement, in the
set $\Im$, is  the following  fundamental fact. The non-extensive
nature of Tsallis entropy would  make sense only in a world where
it were impossible to create uncorrelated systems, and
consequently, a vanishing entanglement. In fact, the Tsallis
entropy is not the sum of the entropies of the parts, not even
when these parts are uncorrelated. The condition that we set, of
non vanishing  entanglement, as earlier said, is equivalent to
ruling out the occurrence of the  splitting of the system into two
uncorrelated parts. Therefore, it corresponds to an ideal
condition for the application of the Tsallis entropy. We shall
come back to this important issue in Section IV.

\subsection{Parametrization of the eigenvalues of
the statistical density matrix} Here we adopt a perspective of
thermodynamic kind, inspired to the lines proposed by Plenio and
Vedral\cite{plenio}. We assume that a purification protocol yields
an entanglement change, $\Delta E_{F}$. We aim at establishing the
change of the non-extensive entropy corresponding to the same
"thermodynamic" transformation. To solve this delicate problem we
imagine the statistical density matrix $\rho$, belonging to the
set $\Im$, to be a function of a real parameter $\xi$, belonging
to an interval $\left[\xi_{1},\xi_{2}\right]$, which can be
thought of as playing the same role as that of variables like
pressure, temperature and volume in the state transformations in
ordinary thermodynamics. We assume that the initial and final
conditions correspond to $\rho\left(\xi_{1}\right)$ and
$\rho\left(\xi_{2}\right)$, respectively, so that $\Delta E_{F}
\equiv E_{F}\left(\rho\left(\xi_{2}\right)\right) -
E_{F}\left(\rho\left(\xi_{1}\right)\right)$. We set the dependence
of $\rho$ on $\xi$ in such a way as to fulfill the constraint
imposed by the norm conservation $ Tr \rho=1$, and among the
infinitely large number of possibilities fitting these conditions,
we select the form most convenient for the purpose of evaluating
$\Delta S_{q} \equiv S_{q}\left(\rho\left(\xi_{2}\right)\right)
-S_{q}\left(\rho\left(\xi_{1}\right)\right)$. Let $P_{m}$ be the
largest of the four eigenvalues of the statistical density matrix
$\rho$ and let us denote with $P_{1}, P_{2}$ and $P_{3}$ the other
three eigenvalues. These eigenvalues are assumed to be functions
of the parameter $\xi$ with the following condition: in the whole
interval $\left[\xi_{1},\xi_{2}\right]$ we have
\begin{equation}
  P_{m}\left(\xi\right)>\frac{1}{2}
\label{positive entanglement}.
\end{equation}
The derivatives $dP_{m}/d\xi$ and $dP_{j}/d\xi$ with $j$ ranging
from $1$ to $3$, are assumed to be finite. The norm conservation
constraint enforces the condition
\begin{equation}
\label{filippo}
\frac{dP_{m}}{d\xi}=-\frac{dP_{1}}{d\xi}-\frac{dP_{2}}{d\xi}
-\frac{dP_{3}}{d\xi}.
\end{equation}

 As earlier pointed out,
the \emph{concurrence} is equivalent to the entanglement of
formation and, in the set $\Im$, the concurrence is a strictly
increasing monotonic function of $P_{m}$. Thus, the change of
$P_{m}$ will be either positive or negative according to whether
the entanglement change is positive or negative.

We shall assume that in the whole interval
$\left[\xi_{1},\xi_{2}\right]$ the derivative $dP_{m}/d\xi$ is
either always positive or always negative. As earlier stressed,
this convenient choice is legitimated by the fact that, as we do
in ordinary thermodynamics, we are considering a state
transformation, and consequently both the entanglement and entropy
changes depend only on the initial and final states, and are
independent of the paths used to connect these two states.

\subsection{Failure of the von Neumann entropy}
Here we show that the von Neumann entropy, namely, the quantum
version of the Shannon information, turns out to be inadequate as
an entanglement indicator. It is so because, as we hereby see, the
von Neumann entropy can either increase or decrease, in
correspondence of the entanglement change, $\Delta E_{F}$,
regardless of whether the entanglement change is positive or
negative. This conclusion agrees with the remarks of the recent
work by Brukner and Zeilinger\cite{attempt5}.

We note that to prove the inadequacy of the von Neumann's entropy as
an indicator of entanglement it is enough to find a case where the
sign of the entropy change is not strictly determined by that of the
entanglement change.
Thus, let us consider the special physical condition corresponding to:
\begin{equation}
     \label{arbitrarychoice}
\frac {dP_{3}}{d\xi}=0
\end{equation}
in the whole interval $\left[\xi_{1},\xi_{2}\right]$. The von
Neumann's entropy, corresponding to $q = 1$, and consequently
denoted as $S_{1}$, reads
\begin{equation}
S_{1}\left( \rho
\right)=-P_{m}\ln P_{m}-\sum_{j=1}^{3} P_{j}\ln P_{j}
\end{equation} and its dependence upon the parameter $\xi$ is given by:
\begin{equation}
\frac{dS_{1}}{d\xi}=-\frac{dP_{m}}{d\xi} \ln \left(
\frac{P_{m}}{P_{1}}\right)+ \frac{dP_{2}}{d\xi} \ln
\left(\frac{P_{2}}{P_{1}} \right).
\end{equation}
Let us consider, for example, the case $dP_{m}/d\xi < 0$. As
earlier noted, in this case the entanglement of the mixed state is
a strictly decreasing function of $\xi$, and the entanglement
change $\Delta E_{F}$ must be negative. However, using Eq.
(\ref{arbitrarychoice}) and Eq. (\ref{filippo}) we see that the
sign of $dS_{1}/d\xi$ depends on the special values selected for
the parameters $P_{m}, P_{1}, P_{2} $ and for the corresponding
$\xi$-derivatives. Thus, it is possible to realize either the
inequality
\begin{equation}
\frac {dP_{m}}{d\xi}< \frac {dP_{2}}{d\xi} \ln \left( \frac
{P_{1}}{P_{2}} \right) / \ln \left( \frac {P_{m}}{P_{1}} \right),
\end{equation}
which would lead to an entropy increase, or the opposite inequality,
which would yield an entropy decrease. This proves that the sign of the
von Neumann's entropy change is not correlated to that of the
entanglement, and that consequently the von Neumann's entropy is not
an adequate entanglement indicator.

\subsection{Search for a critical entropy index}
       The theoretical developments of this subsection
       show that the failure of the von Neumann's entropy
  is the consequence of the fact that the von Neumann's entropy  means
  $q =1$  and this entropy index is smaller than, or equal to,
  a critical value $Q\left(\rho_{1},\rho_{2}\right)$, where $\rho_{1}$ and
  $\rho_{2}$ denote the initial and the final state, respectively.
  Hereby we show, in fact, that for $q > Q\left(\rho_{1},\rho_{2}\right)$ the
  non-extensive entropy becomes a monotonic function of the
  entanglement strength, inversely proportional to it.

  For this purpose, let us study the $\xi$-derivative of the non-extensive
  entropy. This quantity reads:
\begin{equation}
\frac {dS_{q}}{d\xi}= - \frac {q P_{m}^{q-1}}{q-1}
\frac{dP_{m}}{d\xi} \left\{ 1+\sum _{j=1}^{3} \left(
\frac{dP_{m}}{d\xi}  \right)^{-1}\left( \frac{P_{j}}{P_{m}}
\right)^{q-1}  \frac{dP_{j}}{d\xi} \right\}.
\end{equation}
We remind the reader that $P_{m}$ is the largest eigenvalue of
$\rho$, thereby implying the property
\begin{equation}
0 \leq \frac {P_{j}}{P_{m}}<1.
\end{equation}
By using these inequalities and the assumption made in  Section
III B that the derivatives $dP_{j}/d\xi$ and $dP_{m}/d\xi$ are
finite, we obtain
\begin{equation}
\label{limit} \lim _{q \rightarrow +\infty}\left[ 1+\sum_{j=1}^{3}
\left(\frac{P_{j}}{P_{m}}\right)^{q-1}
\left(\frac{dP_{m}}{d\xi}\right)^{-1}
\left(\frac{dP_{j}}{d\xi}\right) \right] = 1.
\end{equation}
This simple result implies that \emph{great enough} values of the
entropy index yield the important property
\begin{equation}
sign\left(\frac{dE_{F}}{d\xi}\right)=sign\left(\frac{dP_{m}}{d\xi}
\right)=-sign\left(\frac{dS_{q}}{d\xi} \right),
\end{equation}
which, in turn, means that increasing entanglement yields a
decreasing entropy, and vice-versa.

At this stage of our path towards the detection of $Q
\left(\rho_{1},\rho_{2}\right)$ we explore two distinct cases
concerning the behavior of $E_{F}\left(\xi\right)$ in the whole
interval $\left[\xi_{1},\xi_{2}\right]$: $(a)$ $dE_{F}/d\xi<0$;
$(b)$ $dE_{F}/d\xi>0$. As earlier mentioned, the arbitrary choice
of a $\xi$-derivative, either always positive or always negative,
is legitimate since the entanglement and the entropy changes are
independent of the path connecting the initial to the final state.
In cases $(a)$ and $(b)$ we shall identify the important auxiliary
functions $q^{\star}\left(\xi\right)$ and
$q^{\star\star}\left(\xi\right)$, respectively. These auxiliary
functions will be used first to build up the functions
$Q^{\star}\left(\xi_{1},\xi_{2}\right)$ and
$Q^{\star\star}\left(\xi_{1},\xi_{2}\right)$, and finally the key
quantity $Q\left(\rho_{1},\rho_{2}\right)$.

  Let us consider
  case $(a)$ first. As a consequence of $dE_{F}/d\xi<0$,
  from the limit  of Eq. (\ref{limit}) we naturally obtain
that a critical value $q^{\star}(\xi)$ exists such that $q >
q^{\star}(\xi)$ yields
\begin{equation}
1+ \sum_{j=1}^{3}\left(\frac{dP_{m}}{d\xi}\right)^{-1}
\left(\frac{P_{j}}{P_{m}}\right)^{q-1}
\left(\frac{dP_{j}}{d\xi}\right)
> 0. \label{E Sq}
\end{equation}
The critical value $q^{\star}(\xi)$ fulfilling the condition of
Eq. (\ref{E Sq}) is not unique. We therefore adopt a criterion to
estimate one of the possible critical values. This will imply that
the resulting $Q(\rho_{1},\rho_{2})$ is not unique either, but, as
shown later, we shall be able to find at least one of them
fulfilling the earlier mentioned properties of
$Q(\rho_{1},\rho_{2})$. The choice that we adopt to find one of
the possible $q^{\star}(\xi)$'s is as follows. We set the
inequality
\begin{equation}
  \left| \frac{dP_{m}}{d\xi}\right| ^{-1}
  \left(\frac{P_{j}}{P_{m}}\right)^{q-1}
    \frac{dP_{j}}{d\xi} <
      \frac{1}{3}, \label{azzurra}
\end{equation}
with the subscript $j$ running from $1$ to $3$. We assume that
this property holds true for any $q > q^{\star}\left(\xi\right)$.
This set of conditions, after an easy algebra, yields
\begin{equation}
 q^{\star} \left(\xi\right) =
 \max\left\{1,\alpha^{\star}_{1}\left(\xi\right),
 \alpha^{\star}_{2}\left(\xi\right),\alpha^{\star}_{3}\left(\xi\right)
 \right\}.
 \label{q star}
 \end{equation}
 As to the definition of $\alpha_{j}^{\star}\left(\xi\right)$, with the
 subscript $j$ running from $1$ to $3$, we must distinguish two
 cases. The first is the case when the constraints
 \begin{equation}
P_{j}\left(\xi\right)>0 \hspace{1em} $and$ \hspace{1em}
dP_{j}/d\xi> 0
 \label{jconstraints1}
 \end{equation}
 hold true. In this case, we set
\begin{equation}
\alpha^{\star}_{j}\equiv 1 +
    \left( \ln\left(\frac{P_{m}}{P_{j}}\right) \right)^{-1}
    \ln\left(3\left| \frac{dP_{m}}{d\xi}\right| ^{-1}
    \frac{dP_{j}}{d\xi}\right).
 \end{equation}
If the constraints of Eq. (\ref{jconstraints1}) do not apply, we
set $\alpha^{\star}_{j}\left(\xi\right)=1$. This implies that for
any $q
> q^{\star}\left(\xi\right)$ the condition
\begin{equation}
     \label{inequalityforE-}
\frac{d}{d\xi} E_{F}\left(\xi\right)<0
\end{equation}
yields
\begin{equation}
\frac{d}{d\xi} S_{q}\left(\xi\right) > 0.
\end{equation}
On the basis of this result we define now the function
$Q^{\star}\left(\xi_{1},\xi_{2}\right)$ as follows:
\begin{equation}
Q ^{\star} \left(\xi_{1},\xi_{2}\right) \equiv \sup_{\xi \in
\left[\xi_{1},\xi_{2}\right] }\left\{{q^{\star}(\xi)}\right\}.
\label{Q star}
\end{equation}
Using Eq. (\ref{inequalityforE-}) and the ensuing inequality for
$dS_{q}/d\xi$ we conclude immediately that for any $q$ fulfilling
the inequality $q > Q^{\star}\left(\xi_{1},\xi_{2}\right)$ the
condition $\Delta E_{F}< 0$ yields $\Delta S_{q} > 0$. In fact, in
this case $\Delta E_{F}$ and $\Delta S_{q} $ can be written under
the form of integrals in the interval
$\left[\xi_{1},\xi_{2}\right]$ with integrands always negative and
positive, respectively.

In case $(b)$ we adopt the same procedure which yields, in this case,
\begin{equation}
 q^{\star\star} \left(\xi\right) \equiv
 \max\left\{1,\alpha^{\star\star}_{1}\left(\xi\right),
 \alpha^{\star\star}_{2}\left(\xi\right),\alpha^{\star\star}_{3}\left(\xi\right)
 \right\}.
 \label{q starstartrue}
 \end{equation}
As to the term $\alpha^{\star\star}_{j}\left(\xi\right)$, in the
case where the constraints
\begin{equation}
P_{j}(\xi) > 0 \hspace{1em}$and$\hspace{1em} dP_{j}/d\xi < 0
\label{jconstraints2}
\end{equation}
hold true, we set
\begin{equation}
\alpha_{j}^{\star\star}\left(\xi\right)\equiv 1 +
      \left(\ln\left( \frac{P_{m}}{P_{j}}\right)\right)^{-1}
      \ln \left(3\left(\frac{dP_{m}}{d\xi}\right)^{-1}
      \left|\frac{ dP_{j}}{d\xi}\right|\right).
\end{equation}
If the constraints of the Eq. (\ref{jconstraints2}) do not apply
we set $\alpha^{\star\star}_{j}\left(\xi\right)=1$. The
counterpart of Eq. (\ref{Q star}) becomes
\begin{equation}
Q ^{\star\star} \left(\xi_{1},\xi_{2}\right) \equiv \sup_{\xi \in
\left[\xi_{1},\xi_{2}\right] }\left\{q^{\star\star}(\xi)\right\}.
\label{Q starstartrue}
\end{equation}
Of course, the condition $\Delta E_{F} > 0$ yields $\Delta S_{q} <
0$.

Note that we have not discussed the problem of the possible
divergence of $Q ^{\star } \left(\xi_{1},\xi_{2}\right)$ or $Q
^{\star\star} \left(\xi_{1},\xi_{2}\right)$. We shall come back to
this issue in the following section where we will consider a
special parametrization of the eigenvalues, without losing any
generality, within which, as we shall see, this critical index
will be proved to be finite.

\subsection{Searching a critical value of the entropy index, beyond
which the Tsallis entropy exhibits a correct dependence on the entanglement
strength}

Now let us see how to use the earlier results to make predictions
in the case where the transformation, and the ensuing entanglement
change as well, is defined only by the initial and the final
states,  $\rho_{1}$ and $\rho_{2}$, with the density matrix
belonging to the set $\Im$ defined by Eq. (\ref{working set}). The
main idea is to build up auxiliary states, $\rho^{(1)}_{B}$ and
$\rho^{(2)}_{B}$, equivalent to $\rho_{1}$ and $\rho_{2}$ as far
as their entanglement and entropy are concerned, but fulfilling
the condition of being connected the one to the other by the
$\xi$-transformation of Section III C. This makes these states
compatible with the earlier prescriptions, and thus with the
earlier results. These two states are defined as follows:
\begin{equation}
     \label{nameless1}
\rho^{(1)}_{B}\equiv P_{m}^{(1)}|e_{m}\rangle \langle e_{m}|+
\sum_{j=1}^{3}P_{j}^{(1)}|e_{j} \rangle \langle e_{j}|
\end{equation}
and
\begin{equation}
     \label{nameless2}
     \rho^{(2)}_{B}\equiv P_{m}^{(2)}|e_{m}\rangle \langle e_{m}|+
\sum_{j=1}^{3}P_{j}^{(2)}|e_{j} \rangle \langle e_{j}|,
\end{equation}
where the set $\big\{|e_{m}\rangle, |e_{j}\rangle, j=1,2,3\big\}$
is the Bell basis set \cite{bw}, no matter what the order is. The
symbols $P_{j}^{(1)}$ and $P_{m}^{(1)}$ denote the eigenvalues of
the density matrix $\rho_{B}^{(1)}$, and, obviously, the symbols
$P_{j}^{(2)}$ and $P_{m}^{(2)}$ denote the eigenvalues of the
density matrix $\rho_{B}^{(2)}$. These quantum states have the
following properties: (i) They belong to the set $\Im$; (ii)
$E_{F}(\rho_{1})=E_{F}\left(\rho^{(1)}_{B}\right)$,
$E_{F}(\rho_{2})= E_{F}\left(\rho^{(2)}_{B}\right)$; (iii)
$S_{q}(\rho_{1})=S_{q}\left(\rho^{(1)}_{B}\right)$,
$S_{q}(\rho_{2})= S_{q}\left(\rho^{(2)}_{B}\right)$.

Now let us introduce the transformation $\Xi_{\xi}
\left[\rho^{(1)}_{B},\rho^{(2)}_{B}\right ]$ defined by:
\begin{equation}
     \label{transformation}
\Xi_{\xi} \left[\rho^{(1)}_{B},\rho^{(2)}_{B}\right ]
\left(\rho^{(1)}_{B}\right) \equiv
P_{m}\left(\xi\right)|e_{m}\rangle \langle e_{m}|+
\sum_{j=1}^{3}P_{j}(\xi)|e_{j} \rangle \langle e_{j}|,
\end{equation}
where the $\xi$-evolutions of $P_{m}(\xi)$ and $P_{j}(\xi)$ are
given by
\begin{equation}
     \label{transformation1}
P_{m}\left(\xi\right)\equiv P_{m}^{(1)} + \xi
\left(P_{m}^{(2)}-P_{m}^{(1)} \right)
\end{equation}
and
\begin{equation}
     \label{transformation2}
P_{j}\left(\xi\right)\equiv P_{j}^{(1)} + \xi \left(P_{j}^{(2)}-
P_{j}^{(1)} \right),\hspace{1em}
\end{equation}
with $j$ running from $1$ to $3$, respectively, and $\xi \in
[0,1]$. The transformation $\Xi_{\xi}$ has the required
properties: (a) It keeps the state $\Xi_{\xi}
\left[\rho^{(1)}_{B},\rho^{(2)}_{B}\right ]
\left(\rho^{(1)}_{B}\right)$ within the set $\Im$ for every value
of $\xi$ belonging to the interval $[0,1]$; (b)
$\Xi_{0}\left[\rho^{(1)}_{B},\rho^{(2)}_{B}\right] \left(
\rho^{(1)}_{B} \right)= \rho^{(1)}_{B} $; (c)
$\Xi_{1}\left[\rho^{(1)}_{B},\rho^{(2)}_{B}\right] \left(
\rho^{(1)}_{B} \right)= \rho^{(2)}_{B}$. Note that the functions
$P_{m}(\xi), P_{1}(\xi), P_{2}(\xi)$ and $P_{3}(\xi)$, are defined
in the interval $\left[\xi_{1},\xi_{2}\right]$. They fulfill the
properties of Eq. (\ref{positive entanglement}), the parameter
conditions of Section IIB, the relation $dP_{m}/d\xi> 0$, in the
case $P_{m}^{(1)}<P_{m}^{(2)}$, and the relation $dP_{m}/d\xi <0$,
in the case $P_{m}^{(1)}> P_{m}^{(2)}$. This makes it possible for
us to use $Q^{\star}\left(\xi_{1},\xi_{2}\right)$ of Eq. (\ref{Q
star}) and $Q^{\star\star}\left(\xi_{1},\xi_{2}\right)$  of Eq.
(\ref{q starstartrue}), and the relations on which these
quantities rest as well, to derive
$Q\left(\rho_{1},\rho_{2}\right)$. This is done as follows. We
write the explicit forms that
$Q^{\star}\left(\xi_{1},\xi_{2}\right)$ and
$Q^{\star\star}\left(\xi_{1},\xi_{2}\right)$ get when $\xi_{1} =
0$ and $\xi_{2} = 1$. Using the transformation of Eq.
(\ref{transformation1}) and Eq. (\ref{transformation2}), we obtain
for $Q^{\star}(0,1)$ the following expression:
\begin{equation}
     \label{firstQ}
    Q^{\star}(0,1)\equiv \sup_{\xi\in
    [0,1]}\max\left\{1,
    \beta^{\star}_{1}\left(\xi\right),
    \beta^{\star}_{2}\left(\xi\right),
    \beta^{\star}_{3}\left(\xi\right)
    \right\}.
\end{equation}
If the constraints
\begin{equation}
\label{jconstaints3} P^{(2)}_{j}\left(\xi\right) >
P^{(1)}_{j}\left(\xi\right) \hspace{1em}$and$\hspace{1em}
P_{j}^{(1)}+\xi\left(P_{j}^{(2)}-P_{j}^{(1)}\right)>0
\end{equation}
 hold true, we set
\begin{equation}
\beta^{\star}_{j}\left(\xi\right)\equiv
1+\frac{\ln\left(3\frac{P_{j}^{(1)}-P_{j}^{(2)}}
{P_{m}^{(2)}-P_{m}^{(1)}} \right)}{\ln\left(\frac{P_{m}^{(1)}+
\xi\left(P_{m}^{(2)}-P_{m}^{(1)}\right)}{P_{j}^{(1)}+\xi\left(P_{j}^{(2)}-
P_{j}^{(1)}\right)} \right)}.
\end{equation}
If the constraints of Eq. (\ref{jconstaints3}) do not apply,  we
set $\beta^{\star}_{j}\left(\xi\right)=1$.

     Note that this mathematical definition must be interpreted as follows.
     First of all, we consider a given value of $\xi$ belonging to the
     interval $[0,1]$. Then we make the index $j$ run from $1$ to
     $3$. We select the indexes $j$ fulfilling the conditions
     $P^{(2)}_{j}\left(\xi\right) > P^{(1)}_{j}\left(\xi\right)$
     and $P_{j}^{(1)}+\xi\left(P_{j}^{(2)}-P_{j}^{(1)}\right)>0$ and
     calculate $\beta_{j}^{\star}\left(\xi\right)$ using the above
     definitions. Then we take the maximum of the values of a set whose components
     are given by $\beta_{j}^{\star}\left(\xi\right)$ and by $1$. Then, we make $\xi$
     explore all the possible values of the interval $[0,1]$. Thus, we get an
     infinite set of maxima, from which we select the supremum.
     The resulting number defines the critical index
     of the left hand side of Eq. (\ref{firstQ}).

     The resulting critical index is finite. To prove this important
     property we proceed as follows. We note that the term that could
     make $Q^{\star\star}(0,1)$ diverge is
     $\frac{P_{m}^{(1)}+ \xi\left(P_{m}^{(2)}-P_{m}^{(1)}\right)}
     {P_{j}^{(1)}+\xi\left(P_{j}^{(2)}-P_{j}^{(1)}\right)}$.
     We denote this term by $\gamma \left(\xi \right)$.
     The special condition resulting in the divergence of the
     critical index would be given by $\gamma\rightarrow 1^{+}$.
     We observe that $\gamma \left(\xi \right)$ is either an increasing
     (decreasing) or a constant function of $\xi$ depending on whether the
     quantity $P_{j}^{(1)} P_{m}^{(2)}-P_{m}^{(1)} P_{j}^{(2)}$ is positive
     (negative) or equal to $0$. So the minimum value of
     $\gamma \left(\xi \right)$ is
     $\gamma(0)=P_{m}^{(1)}/P_{j}^{(1)}$, in the case of $d\gamma / d\xi >0$,
     and $\gamma(1)=P_{m}^{(2)}/P_{j}^{(2)}$,
     in the case of $d\gamma / d\xi<0$. In the remaining case
     $d\gamma /d\xi=0$, the two minima get the same value. From this properties
     we obtain the following inequality
\begin{equation}
     Q^{\star}(0,1)\leq \max_{j=1,2,3} \left\{1 +
     \left(\ln\left(\min_{j=1,2,3}\left\{\frac{P_{m}^{(1)}}{P_{j}^{(1)}},
     \frac{P_{m}^{(2)}}{P_{j}^{(2)}}\right\}\right)\right)^{-1}
     \left|\ln\left(3\left|\frac{P_{j}^{(2)}-P_{j}^{(1)}}
     {P_{m}^{(1)}-P_{m}^{(2)}}\right|\right)\right|
     \right\},
     \label{finitefirstQ}
     \end{equation}
     proving that $Q^{\star}(0,1)$ is \emph{finite}.

     As to $Q^{\star\star}(0,1)$, we get
\begin{equation}
     \label{secondQ}
    Q^{\star\star}(0,1)\equiv \sup_{\xi\in
    [0,1]}\max\left\{1,
    \beta^{\star\star}_{1}\left(\xi\right),
    \beta^{\star\star}_{2}\left(\xi\right),
    \beta^{\star\star}_{3}\left(\xi\right)
    \right\}.
\end{equation}
If the constraints
\begin{equation}
\label{jconstaints3}
P^{(1)}_{j}\left(\xi\right) >
P^{(2)}_{j}\left(\xi\right)\hspace{1em} $and$
\hspace{1em}P_{j}^{(1)}+\xi\left(P_{j}^{(2)}-P_{j}^{(1)}\right)>0
\end{equation}
 hold true, we set
\begin{equation}
\beta^{\star\star}_{j}\left(\xi\right)\equiv
1+\frac{\ln\left(3\frac{P_{j}^{(1)}-P_{j}^{(2)}}
{P_{m}^{(2)}-P_{m}^{(1)}} \right)}{\ln\left(\frac{P_{m}^{(1)}+
\xi\left(P_{m}^{(2)}-P_{m}^{(1)}\right)}{P_{j}^{(1)}+\xi\left(P_{j}^{(2)}-
P_{j}^{(1)}\right)} \right)}.
\end{equation}
If the constraints of Eq. (\ref{jconstaints3}) do not apply, we
set $\beta^{\star\star}_{j}\left(\xi\right)=1$.

The criterion adopted to define this critical index is the same as
  that earlier illustrated to properly define the critical index
  of Eq. (\ref{firstQ}). Thus, we can prove that $Q^{\star\star}(0,1)$ is \emph{finite}
     adopting a procedure analogous to that used for
     $Q^{\star}(0,1)$.
     In this case we arrive at the following inequality:
\begin{equation}
Q^{\star\star}(0,1)\leq \max_{j=1,2,3} \left\{1 +
     \left(\ln\left(\min_{j=1,2,3}\left\{\frac{P_{m}^{(1)}}{P_{j}^{(1)}},
     \frac{P_{m}^{(2)}}{P_{j}^{(2)}}\right\}\right)\right)^{-1}
     \left|\ln\left(3\left|\frac{P_{j}^{(1)}-P_{j}^{(2)}}
     {P_{m}^{(2)}-P_{m}^{(1)}}\right|\right)\right|
     \right\},
     \label{finitesecondQ}
\end{equation}
which shows in fact that also $Q^{\star\star}(0,1)$ is
\emph{finite}.

     At this stage we can finally define the critical value
     $Q\left(\rho_{1}, \rho_{2}\right)$. This is given by
  \begin{equation}
Q\left(\rho_{1}, \rho_{2}\right) \equiv \max
\left\{Q^{\star}(0,1),Q^{\star\star}(0,1) \right\}.
\end{equation}
On the basis of the earlier described theoretical treatment we can
conclude that $Q\left(\rho_{1}, \rho_{2}\right)$ is \emph{finite}
and that for any initial and final states, $\rho_{1}$ and
$\rho_{2}$, respectively, belonging to the set $\Im$, with
different entanglement, $E_{F}\left(\rho_{1}\right)\neq
E_{F}\left(\rho_{2} \right)$, the corresponding entropy change
$\Delta S_{q}$ is positive or negative, according to whether
$\Delta E < 0$ or $\Delta E > 0$. Note that we have found
$+\infty>Q(\rho_{1}, \rho_{2}) \geq 1$. As a consequence of the
pseudoadditivity of Eq. (\ref{breakdownofthe additiveproperty}),
the adoption of a value of the entropy index larger than the unity
makes the entropy of the whole system smaller than the sum of the
entropies of the two parts. However, in this paper we never make a
direct use of this property, since, as earlier stressed, our
treatment is valid only in the case of non vanishing entanglement,
which rules out the possibility of realizing the factorized
condition behind Eq. (\ref{breakdownofthe additiveproperty}).

\subsection{From the non-extensive entropy to the entanglement of
formation} As a purpose of this subsection, we try to prove a
property that is the reverse of that discussed in Section III E.
Ideally, the reverse of the property of Section III E should be
expressed as follows. Let us focus our attention on a
transformation  from an initial state $\rho_{1}$ to a final state
$\rho_{2}$, both belonging to the  set $\Im$. Let us consider a
case where this transformation makes the non-extensive entropy
$S_{q}$ increase (decrease). Then, the entanglement decreases
(increases) if an entropy index $q$ larger than the critical value
$Q$ is adopted. Unfortunately, we cannot prove this property in
this attractive form, but only under weaker conditions. This is so
because a transformation  resulting in an entropy change does not
necessarily imply an entanglement change.  We notice that the
entanglement, expressed in the set $\Im$, is a function of the
eigenvalue $P_{m}$ only, while the non-extensive entropy is a
function of all four eigenvalues. Thus, the entropy can change
without implying a corresponding entanglement change. The same
difficulty is shared by the non-extensive entropy. However, upon
increase of the entropy index $q$ the dependence of the
non-extensive entropy on the other three eigenvalues becomes
weaker and weaker. In the case of \emph{enough great} values of
the entropy index $q$ the non-extensive entropy becomes virtually
independent of the other three eigenvalues. This is the reason why
in Section III we could find a way to make the non-extensive
entropy become a monotonic function of entanglement. We want to
remark that in general the entropy critical index is not the same
as that used in Section III.

We think that one of the benefits resulting from the adoption of
the set $\Im$, and of very large entropy indices as well, is that
the margin of entanglement dependence on entropy is significantly
reduced. Nevertheless, we are forced to make a weaker request for
the reverse of the property discussed in Section E. We shall show,
in fact, that if the entropy increases (decreases),  \emph{and the
entanglement changes}, then the entanglement decreases
(increases), for entropy indices $q$ larger than a critical value
$Q^{(S)}$, \emph{not necessarily equal to $Q$}. The conditions
emphasized by the adoption of italics make the property weaker
than we would wish. Even in this case we have to assume the
entropy index to be larger than a critical value. We denote this
critical value with the symbol $Q^{(S)}$ because, as earlier
mentioned, we cannot prove that it is identical to the critical
entropy index $Q$ of Section III.

In the case of entropy increase, by expressing the non-extensive
entropy as a function of its four eigenvalues, we get

\begin{equation}
     \label{giveaname}
\left( \frac {P_{m}^{(2)}}{P_{m}^{(1)}}
\right)^{q} < 1+\sum_{j=1}^{3} \left(
\left(\frac{P_{j}^{(1)}}{P_{m}^{(1)}} \right)^{q}-
\left(\frac{P_{j}^{(2)}}{P_{m}^{(1)}} \right)^{q} \right).
\end{equation}
Since the two eigenstates have different entanglement we have
$P_{m}^{(2)}/P_{m}^{(1)} \neq 1$. Since the inequality of Eq.
(\ref{giveaname}) must hold true in the case of entropy indices
arbitrarily larger than $Q^{(S)}$, and consequently, must hold
true also for values much larger than the unity, we reach the
conclusion that $P_{m}^{(2)} < P_{m}^{(1)}$, which in the set
$\Im$ is equivalent to $\Delta E_{F}<0$. The opposite conclusion
would be reached in the case of a negative $\Delta S_{q}$.

In spite of the earlier restrictions, we can use the obtained
results to illustrate one of the most interesting findings of this
work. This is as follows. Let us consider  a generic subset
$\Im'$, of the set $\Im$, only fulfilling the request of
containing a \emph{finite} number of states, with \emph{different}
entanglements. Then, we can conclude that these entanglements are
\emph{equivalent} to the \emph{inverse} of the non-extensive
entropy, provided that entropy indices $q$ are larger than a given
value $Q_{\Im ^{\prime}}$, which is given by the following formula
\begin{equation}
     \label{fundamentalresult}
Q_{\Im ^{\prime}}\equiv \max\left\{Q\left(\rho_{i},
\rho_{j}\right), \forall \rho_{i}, \rho_{j} \in \Im^{\prime},
i\neq j  \right\}.
\end{equation}
In the set $\Im'$  for entropy indices larger than the critical
value the ordering in the direction of increasing (decreasing)
entanglement is equivalent to ordering in the direction of
decreasing (increasing) entropy. A significant consequence of this
is that entropy minimization yields the maximally entangled state
and the entropy maximization the minimally entangled state.

An attractive, albeit heuristic way, of illustrating the same
conclusions is given by the
following formula:
\begin{equation}
E^{(eff)}_{q}\left(\rho \right)\equiv \frac{2}{\pi} \arctan
\left\{\sum_{i=1}^{4} \Theta \left(P_{i}-\sum _{k\neq i} P_{k}
\right) \left(3-4S_{2} \left(\rho \right)-4 \sum _{k\neq i}
P_{k}^{2} \right) S_{q}^{-1} \right\}, \label{q ent}
\end{equation}
which establishes a direct connection between entanglement and
entropy. The quantity $E^{(eff)}_{q}$ is ``equivalent'' to the
entanglement, in the sense that it increases or it decreases upon
increasing or decreasing the entanglement strength. Furthermore,
it is equal to $1$ when the entanglement is $1$ and tends to
vanish with the entanglement measure tending to zero. The key
ingredient of this heuristic formula is the term $\arctan$ and the
factor $\sum_{i=1}^{4} \Theta \left(P_{i}-\sum _{k\neq i} P_{k}
\right) \left(3-4S_{2}\left(\rho \right)-4\sum _{k\neq i}
P_{k}^{2} \right)$. Without the first one, the condition $P_{m}
\rightarrow 1^{-}$ would generate divergencies. Furthermore with
$P_{m} \rightarrow (1/2)^{+}$ the inverse of the entropy would
tend to a minimum, which would be different from $0$, which is the
right value. With the factor $\sum_{i=1}^{4} \Theta
\left(P_{i}-\sum _{k\neq i} P_{k} \right) \left(3-4S_{2}\left(\rho
\right)-4\sum _{k\neq i} P_{k}^{2} \right),$ we get rid of the
divergencies and we do succeed in ensuring that the quantity
$E^{(eff)}_{q}$ tends to vanish with the entanglement tending to
zero. Note that this \emph{ad hoc} factor is nothing but the
square of the concurrence. In principle one could express the
concurrence in terms of $S_{2} $, but this would not afford the
attractive condition of the entanglement being a monotonical
increasing function of the inverse of the non-extensive entropy.

In conclusion, we find that entanglement increase implies entropy
decrease and vice versa. This property must be compared with the
results of the work of Abe and Rajagopal\cite{attempt2}. These
authors adopt the principle of entropy maximization under suitable
constraints to infer a plausible form of physical state, and
conclude that the entangled states are the important result of
this maximization process. Here we adopt a different perspective,
based on the fact that the definition of entanglement of formation
is already inspired to statistical mechanics\cite{wootters}.
Within this perspective the state of maximum entanglement
corresponds to the minimum amount of information necessary to
describe the state. Within this same perspective, the amount of
information necessary to describe the state becomes increasingly
larger upon reducing the entanglement strength. From an intuitive
point of view, the occurrence of de-coherence, that is judged by
many authors\cite{zeh} to be the key condition to derive classical
from quantum physics, implies a significant entropy increase.
However, de-coherence, as a form of real wave function
collapse\cite{tessieri}, implies the breakdown, in the long-time
limit, of the entanglement condition, and, as a consequence, the
breakdown of the theory itself of the present paper. The result of
this paper has to be considered within this perspective. As it
appears from the literature on this new and exciting subject, the
thermodynamic significance of the processes of quantum
teleportation is a very delicate and difficult issue.  We are
inclined to believe that the adoption of a non-extensive form of
entropy might be of some relevance, under specific restrictions.
The first is that real wave function collapses are ignored, and
the second is that, in a world dominated by quantum entanglement,
the condition of maximum entanglement is perceived as that
requiring the minimum amount of information. In other words,
increasing entanglement means smaller, rather than  larger,
entropy values.

\section{The Tsallis entropy at work: Dephasing processes in the Bell
basis set}

Before ending this paper, it is convenient to illustrate another
interesting result that does not require a restriction to the
set $\Im$. This has to do with an important result obtained by
Bennett \emph{et al}\cite{bw}. These authors studied the entanglement
changes as a function of a dephasing process. More precisely, they
focused their attention on the transformation
\begin{equation}
     \label{transformation3}
D_{B}=\frac{1}{4}\sum_{i=0}^{3} U^{\dag}_{i}\rho U_{i},
\end{equation}
which brings the initial condition described by the density matrix
$\rho$, expressed in the Bell basis, into the diagonal form
\begin{equation}
     \label{diagonal}
\left[D_{B}\right]_{ij}\equiv \delta_{ij}\left[\rho\right]_{ij},
\end{equation}
where the operators $U_{i}, i=0, 1,2,3$ are respectively $I,
B_{x}B_{x}, B_{y}B_{y}, B_{z}B_{z}$ and $B_{i}$ is the bilateral
rotation of $\pi/2$ around the $i-th$ axis of the space
$S_{1/2}^{(1)}\times S_{1/2}^{(2)}$. This bilateral rotation is
defined by these authors\cite{bw} as
\begin{equation}
B_{i}=\frac{1}{2}\left(I_{2\times 2} -i \sigma_{1i}\right)\times
\left(I_{2\times 2} -i \sigma_{2i}\right).
\end{equation}
Note that the matrix $D_{B}$ of Eq. (\ref{diagonal}) is the
"diagonal" of the statistical density matrix $\rho$ expressed in
the Bell basis and that it results from a random application of
four local unitary transformations, so that moving from the
initial state $\rho^{(1)}$ to the state described by $D_{B}^{(1)}$
the entanglement cannot increase\cite{Horodecki2}. Consequently,
we have
\begin{equation}
 E_{F}\left(\rho^{(1)} \right)\geq E_{F}\left(D_{B}^{(1)}\right).
 \label{E deph}
\end{equation}

We shall analyze these theoretical results by means of the
non-extensive entropy. The first analysis is made by focusing our
attention on the natural values $n > 1$ of the entropy index $q$.
In this special case the non-extensive entropy reads as follows
\begin{equation}
     \label{naturalvalues}
S_{n} \left(\rho\right)= \frac{1-Tr \left(U \cdot \rho \cdot
U^{\dag} \right)^{n}}{n-1}= Tr \frac{\rho -\rho ^{n}}{n-1}.
\end{equation}
Let us define the auxiliary function
\begin{equation} g_{n}\left(x \right)\equiv
\frac{x-x^{n}}{n-1}.
\end{equation}
We note that this is a concave function. On the other hand, several years
ago Wehrl\cite{Wehrl} noticed that in that case we can write
\begin{equation}
     \label{wherl}
     Tr g_{n}\left(D_{B}\right) \geq Tr g_{n}\left(\rho\right).
     \end{equation}
We notice that $S_{n}(\rho) = Tr g_{n}\left(\rho\right)$ and
$S_{n}\left(D_{B}\right) = Tr g_{n}\left(D_{B}\right)$.
Consequently, we can write:
\begin{equation}
     \label{importantinequality}
   S_{n}\left(D_{B}\right) \geq  S_{n}\left(\rho\right)
   \end{equation}

   The results of the dephasing process makes it possible for us to
   generalize the results of Section III. Let us consider
   a transformation from an initial state described by a generic density
   matrix. As to the final state, we set the condition that it belongs
   to the set $\Im$. Let
$p^{(1)}_{m}$ be the maximum of the diagonal elements of the
initial state $\rho^{(1)}$, expressed in the Bell basis set. Let
us suppose also that $p^{(1)}_{m}$ is larger than the maximum
eigenvalue of the density matrix $\rho^{(2)}$, referring to the
final state. This condition is expressed by the relation:
\begin{equation}
p^{(1)}_{m}\equiv
\max_{i=1,2,3,4}\left[\rho^{(1)}\right]_{ii}>P_{m}^{(2)}>\frac{1}{2}.
\label{deph cond}
\end{equation}
As a consequence of this relation we
have
\begin{equation}
E_{F}\left(\rho^{(1)}\right)\geq E_{F}\left(D^{(1)}_{B}\right)>
E_{F}\left(\rho^{(2)}\right)>0.
\end{equation}
This is a transformation with a decreasing entanglement. On the
basis of the results of Section III and of Eq.
(\ref{importantinequality}), we are in a position to find values
of the entropy index $q$ such that the non-extensive entropy of
the final state is larger than that of the initial state. This is
done as follows. We move from the initial condition $\rho^{(1)}$
to $D^{(1)}_{B}$, through the dephasing process earlier described.
As we have seen, with the adoption of natural values, larger than
the unity, for the entropy indices, the entropy does not decrease.
This means
\begin{equation}
     S_{n}\left(D^{(1)}_{B}\right)\geq
S_{n}\left(\rho^{(1)} \right).
\end{equation}
According to our assumptions, $D_{B}^{(1)}$ and $\rho^{(2)}$
belong to the set $\Im$. Thus, we know, on the basis of the
results of Section III, that there exists a critical value of the
entropy index, $Q \left (D_{B}^{(1)},\rho^{(2)}\right)$, beyond
which the non-extensive entropy increases. If we choose critical
values of the entropy index that are natural numbers larger than
\begin{equation}
N \equiv \left[ Q\left(D_{B}^{(1)},\rho^{(2)} \right)\right],
\label{natural critical value}
\end{equation}
 we conclude that the non-extensive entropy increases.
      As earlier anticipated, this has the effect of
making more general the results of Section III.

\section{conclusions}

This paper shows that in the set $\Im$, enforcing the important
condition of a non vanishing entanglement, the Tsallis entropy is
a monotonic and decreasing function of the increasing
entanglement. The entanglement is, in turn, a monotonic and
decreasing function of the increasing entropy under the key
restriction of transformations yielding an entanglement change.
This conclusion  was reached adopting a perspective taking the
warning of a recent paper by Horodecki \emph{et
al}\cite{Horodecki2} into account. As a matter of fact, these
authors show that the principle of entropy maximization yields
fake entanglement, and consequently becomes questionable. We share
the conviction of these authors and adopt in fact an approach that
does not rest on the Jaynes principle\cite{jaynes1,jaynes2}. Thus
we establish a comparison between entanglement and non-extensive
entropy without invoking the Jaynes principle. We do not need to
maximize entropy after minimizing entanglement, as they
do\cite{Horodecki2}, and the monotonic dependence of entropy on
entanglement is a natural consequence of the adoption of suitably
large entropy indices.

This means that we share the view of Rajagopal and
Abe\cite{attempt2} that a non-extensive form of entropy can prove
to be a convenient tool to study quantum teleportation. In this
sense, this paper contributes deepening our understanding about
the significance of the Tsallis entropy. This entropy indicator
does not split into the sum of two independent contributions, when
applied to a system consisting of two uncorrelated subsystems.
This suggests that this kind of entropy might be a proper
theoretical tool only when applied to cases where the re-partition
into two uncorrelated systems is impossible. Quantum mechanical
systems, in principle, are significant examples of where this
condition applies, if environmental de-coherence, or other kind of
de-coherence processes, are ignored. In this condition the Tsallis
entropy, according to the main result of this paper, seems to work
properly, provided that the warning of Ref.\cite{Horodecki2} is
taken into account. This is where our procedure departs from the
point of view of Rajagopal and Abe\cite{attempt2}. Their approach
is still based on the Jaynes principle, supplemented by the choice
of a suitable additional constraint, concerning the fluctuations
around the average, as well as the ordinary constraint on the mean
value (see also Ref.\cite{rajagopalalone}). This procedure yields
convincing, although non general conclusions. Our approach, which
unfortunately shares the lack of generality of
Ref.\cite{attempt2}, is based on a different perspective, aiming
at identifying the inverse of entanglement with the non-extensive
entropy.

We think that the alternative perspective adopted in the present
paper might contribute, as Refs.\cite{attempt2,rajagopalalone} do,
to a better understanding of the thermodynamic nature of
entanglement. We are afraid that the non-extensive entropy might
become inefficient when we leave the physical condition where the
no-cloning theorem and the principle of no-increasing
entanglement, recently found by Horodecki and
Horodecki\cite{thetwoh}, is broken. According to these authors the
occurrence of real wave function collapses, incompatible with the
restriction of adopting unitary transformations, provokes the
breakdown of this equivalence. In our opinion, the occurrence of
real wave function collapses is incompatible with the restriction
of working on the set $\Im$, which enforces the condition of a non
vanishing entanglement. Thus, we expect that in that case the
theory of this paper, and with it the non-extensive entropy, does
not work.  To explore the uncertain border between quantum and
classical mechanics we probably need to adopt a still more
advanced perspective.

\end{document}